\documentclass[aip,pof,reprint,superscriptaddress]{revtex4-1}
\pdfoutput=1
\usepackage[pdftex,bookmarks,colorlinks]{hyperref}
\usepackage[pdftex]{graphicx}
\usepackage{natbib}
\usepackage{amssymb,amsmath}
\usepackage{color}
\usepackage{longtable}
\DeclareGraphicsExtensions{.pdf,.png,.jpg}
\include{electrosprayer.bib}

\begin{document}



\title{Surface tension effects on immersed electrosprays}


\author{\'Alvaro G. Mar\'in}
\email{a.marin@unibw.de}
\affiliation{Bundeswehr Universit\"at M\"unchen, Neubiberg.}%
\author{Ignacio G. Locertales}
\affiliation{Escuela de Ingenieros Industriales, Universidad de Malaga.}
\author{Antonio Barrero}
\affiliation{Escuela de Superios de Ingenieros, Universidad de Sevilla.}

\begin{abstract}

Electrosprays are a powerful technique to generate charged micro/nanodroplets. In the last century, the technique received extensive study and successful applications, including a Nobel price in  Chemistry. However, nowadays its use in microfluidic devices is still limited mainly due to a lack of knowledge of the phenomenon when the dispersing fluid is immersed in another inmiscible liquid. The  ``immersed electrosprays'' share almost identical properties as their counterparts in air. Things however change when surface active agents are added to the host liquid, which are normally used in lab-on-chip applications to stabilize the generated emulsions. In this work, we review the main properties of the immersed electrosprays in liquid baths with no surfactant, and we methodically study the behavior of the system for increasing surfactant concentrations. The different regimes found are then analyzed and compared with both classical and more recent experimental, theoretical and numerical studies. A very rich phenomenology is found when the surface tension is allowed to vary in the system. More concretely, the lower states of electrification achieved with the reduced surface tension regimes might be of interest in biological or biomedical applications in which the free charge is normally hazardous for the encapsulated entities.  

\end{abstract}

\maketitle


\section{Introduction}
The generation of emulsion droplets of controllable size and structure motivates hundreds of publications in the literature and drags attention from very different types of industries: optics\cite{Drzaic}, cosmetics \cite{cosmetics} or biomedicine\cite{dropletreview}. MMES and Microfluidic technology\cite{whitesides} permit high control on droplet size and easy post-processing on the produced particles \cite{Barr:rev}. Nonetheless, most of the existing techniques based on these microfluidic technologies for particle production have basically two serious constraints: one of them is the low production rate and the other the limitation in particle diameter. The last constraint can be overcome employing techniques on which the final droplet size does not depend on the channel or nozzle sizes. Some approaches at this respect can be found on Anna and Mayer \cite{annamayer}, who employed surfactants at critical concentrations on a microfluidic flow-focusing device to find regimes similar to the so-called tip-streaming. On a different approach, Suryo and Basaran \cite{suryobasaran} predicted numerically a regime with the same qualities as tip-streaming in surfactant-free co-flowing devices under special conditions; Mar\'in, del Campo and Gordillo \cite{Maringordi} recently confirmed it experimentally and demonstrated that the regime is able to generate particles down to the micron size.
\begin{figure}[hbp!]
\centering

\includegraphics[width=0.7\columnwidth]{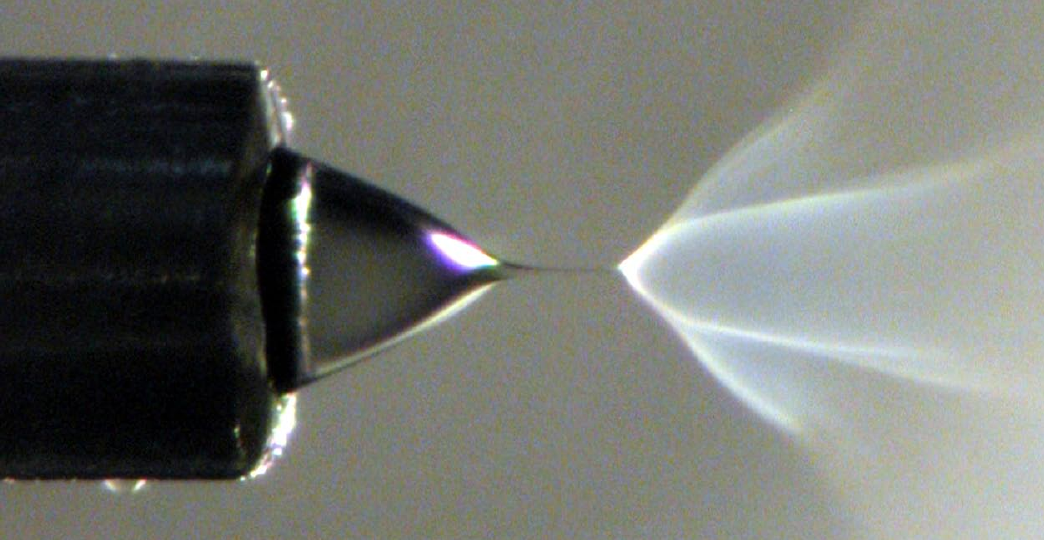}

\caption{Typical aspect of an immersed electrospray in an inviscid dielectric liquid: Metallic capillary, Taylor cone, micro-liquid jet and hydrosol with its characteristic plumed-shape\label{fig:es}. The picture corresponds to sample G4; the outer needle diameter is 0.8mm.}

\end{figure}

Among those techniques able to reach submicrometric sizes, the electrospray is probably one of the most popular due to its spread use in mass spectrometry, application which was honored in 2002 with a shared Noble Prize of Chemistry to Prof. John Fenn. Mass spectrometry has been indeed the only successful application that the electrospray has found so far. Nonetheless, there is still an important number of publications every year suggesting interesting new applications in very different fields, ranging from biomedical to electronics applications\cite{almeria2011multiplexed,deng2011electrospray,gomez2011fundamentals,lozano2011electrospray}. 

In a typical electrospray configuration a conductive liquid is pumped at certain flow rate through a tube, and a meniscus will be formed at its end. The liquid is connected to one of the poles of a high voltage power supply and a counter electrode is placed at some distance from the exit of the tube. Due to the electrical field produced, the free charge in the conductive liquid meniscus will tend to relax towards the liquid surface, generating a net normal electric stress that opposes surface tension. As G. I. Taylor explained in on of his pioneering works, an interesting equilibrium is then reached at a certain voltage between the two surface forces, which obliges the meniscus to adopt a conical shape \cite{taylor1964disintegration}. This equilibrium would only be stationary if the charges were frozen at the interface, which would only happen ideally for liquids with infinite conductivity. In contrast, liquids with finite conductivity manifest electrical shear stresses that lead to the formation of a liquid jet close to the cone apex which breaks eventually downstream intro drops due to capillary instabilities, giving rise to an aerosol of charged droplets (see figure \ref{fig:es}). The jet or droplet diameter can vary substantially depending mainly on the liquid conductivity, ranging from hundreds of micrometers for the least conducting liquids to a few nanometers for the most conducting ones. Finer control on the jet diameters can be achieved by manipulating the injected flow rate \cite{Loscermora} or the voltage \cite{gundabala2010current,larriba2011production} depending on the regime in which the electrospray is manipulated.

Additionally the technique permits great flexibility in the applications, e.g. it also permits encapsulation of substances by the combination of two liquids into the so-called ``coaxial-jet electrospray'' \cite{loscertales2002science}. The technique has proved to be extremely efficient encapsulating substances, not only as spherical micro-capsules, but also in the shape of micro/nano-fibers\cite{diaz2006encapsulation,diaz2006controlled,loscertales2004electrically}. An interesting perspective would be to implement such a powerful atomization technique into a microfluidic device to benefit from the enhanced control of such systems. Barrero et al. \cite{Barr:2003} made the first step towards such objective demonstrating that electrosprays operated in liquid dielectric baths (from now on ``immersed electrosprays'') have analogue properties as their counterparts in air. 
Since then, some have attempted to implement the technique in microfluidic devices \cite{KimH,japos}, with positive but only exploratory results so far. Unfortunately, the implementation of electrohydrodynamic systems in microfluidics has found problems related with the control of the generated charged hydrosol. The charged micro/nanodroplets tend to expand and repeal each other within the whole volume of the microdevice, and due to the confinement of the system, they end up colliding and accumulating at the micro device walls. This produces disturbances in the electrical field surrounding the Taylor cone and has crucial consequences to its stability. Gundabala et al.\cite{gundabala2010current, ecoflow} solved such situation by implementing the electrospray in a cylindrical co-flow system consisting of concentric capillaries. The immersed electroprays are however extremely sensitive to the interfacial tension and to the hydrodynamic shear stress that the external medium applies to the Taylor cone which might compete against the natural electrical shear stresses applied at the Taylor cone interface. These properties have been only marginally treated in the literature and there is therefore a lack of understanding on their effects.

A large fraction of the current development on microfluidics is devoted to the formation of emulsions of controllable size and structure. Emulsions are energetically metastable and need of stabilizing agents as surfactants. Although their use is widespread in daily life and in laboratories, very often its dynamics are complex and unknown, but their use has an impact on biochemical assays\cite{Baret:2012} as well as in systems as those reported here. The electrified meniscus in an electrospray maintains a delicate equilibrium of normal and shear stresses at its interface, and therefore the presence of surfactants can be extremely harmful for its stability. The aim of the present work is therefore to study in detail the effect of surfactants and be able to give some rule of thumb when choosing the experimental parameters for an immersed electrospray to work according to the needs of the user. 

To achieve these objectives, we performed electrical current and droplet measurements for different liquids, surfactant concentrations and flow rates in a non-confined device. Such measurements will yield information about the regimes in which the electrospray is being run, and therefore about mass and charge transport mechanisms within the liquid. With such information we will be able to make a proper comparison with the standard electroatomization in air and to draw conclusions on which regimes would be appropriate for confined experiments in microfluidics or other different applications.

The article is organized as follows: section \ref{sec:scale} reviews the classical scaling laws of electrospraying in air. In section \ref{sec:setup} we give all experimental details of the experimental set-up, liquids employed, and measurement techniques. Section \ref{sec:clean} reviews results immersed electrosprays in clean baths (without surfactants). We continue with baths with surfactants in section \ref{sec:surfactant}, which we separate in two extreme cases: high (subsection \ref{sec:high}) and low conductivity dispersed liquids (subsection \ref{sec:low}). We give a brief description of the phenomenology when inverted viscosity ratios are used in section \ref{sec:inverted}, and end up with summary and final conclusions in section \ref{sec:summary}.

\section{Scaling Laws in Electrosprays\label{sec:scale}}

The mass and charge transport mechanisms in the electrospray manifest experimentally in scaling laws for the electrical current and the droplet size versus the flow rate. In the last 20 years, many have explored different electrospray regimes, with different scalings and properties\cite{MoraReview}. However, it is generally accepted that the regime with the most steady and robust performance, and which yields the most monodisperse aerosols, is the so-called \emph{cone-jet} regime\cite{Cloupeau,gomez2011fundamentals}. 
Recent numerical simulations have confirmed some of these scaling laws using different approaches\cite{higuera2004current,higuera2005electrosprays,higuera2010electrodispersion,higuera2010numerical,BasaranNature}. The current study will not discuss details of the scaling laws found, but will only use those most widely accepted to probe intro the physics of the immersed electrospray. 

Within the cone-jet regime \cite{Cloupeau}, we can find scaling laws to relate the input parameters of the system, basically the flow rate $Q$ and the applied voltage drop $V$, with the output parameters, i.e. the electrical current transported by the liquid $I$ and the droplets size $d$
The scaling laws to be used as a reference are inspired by the pioneering work of Loscertales and de la Mora \cite{Loscermora} and by Ga\~nan et al.\cite{Ganan97}. When high conductivity liquids with intermediate viscosities are operated at small flow rates in a gaseous medium, one can find the following scaling laws, expressed in dimensionless form:

\begin{equation}  \label{eq:Ilaw}
\frac{I}{I_{o}}=a \left(\frac{Q}{Q_{o}}\right)^{1/2},
\end{equation}

\begin{equation}  \label{eq:dlaw}
\frac{d}{d_{o}}=b \left(\frac{Q}{Q_{o}}\right)^{1/3}.
\end{equation}
Where we have defined the characteristic values of flow rate $Q_o$, electrical current $I_o$ and droplet diameter $d_o$ using the density $\rho$, conductivity $K$ and surface tension $\gamma$ of the conducting liquid as follows:

\begin{equation}\label{eq:Qo}
Q_{o}=\frac{\epsilon_{o} \gamma}{\rho K},
\end{equation}

\begin{equation}\label{eq:Io}
I_o=\gamma \sqrt{\frac{\epsilon_o}{\rho}},
\end{equation}

\begin{equation}\label{eq:do}
d_o=\left(\frac{\gamma \epsilon_o^2}{\rho K^2}\right)^{1/3}.
\end{equation}
Different other scalings with different exponents can be found for different range of flow rates, conductivities, viscosity and, as we will see in the following, of the interfacial tension. The most relevant ones will be discussed further below. 

\begin{figure}[hbp!]
\centering

\includegraphics[width=0.6\columnwidth]{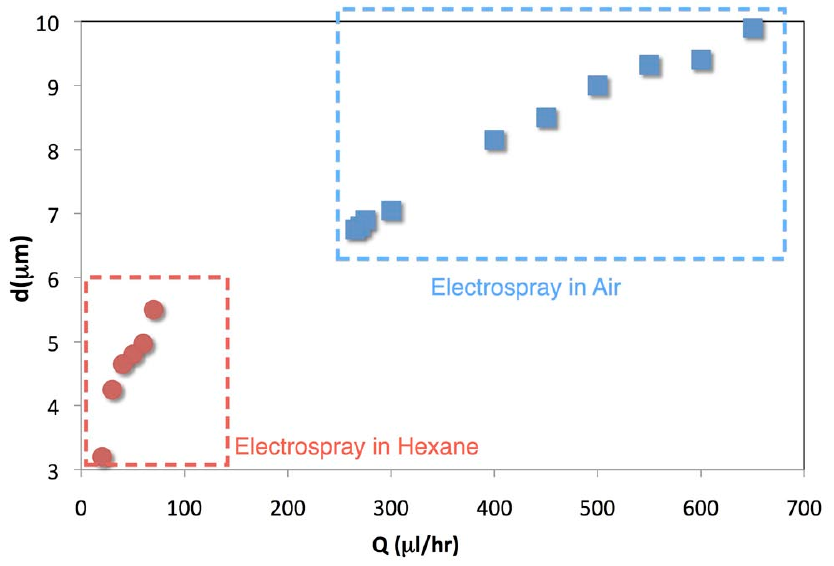}

\caption{Comparison of droplet size distributions: glycerol droplets in air vs glycerol droplets in hexane \label{fig:airvshexane}}

\end{figure}

To outline some of the advantages of the immersed electrosprays, we compare the droplet diameters obtained with an immersed electrospray with the same liquid electrosprayed in air. The sample chosen is G5 since its characteristic diameter fits well with the range of measurement in the apparatus. As commented above, when dispersing a conducting liquid into a dielectric in the cone-jet regime, the characteristic flow rate decreases according mainly to the decrease in surface tension. Consequently, the droplet diameter decreases notably as can be clearly seen in figure \ref{fig:airvshexane}. Operating the electrospray in a dielectric liquid permits us to decrease the flow rate in almost one order of magnitude, with the corresponding decrease in droplet size.

\section{Experimental set-up and methods}\label{sec:setup}

\begin{figure}
\centering

\includegraphics[scale=0.6]{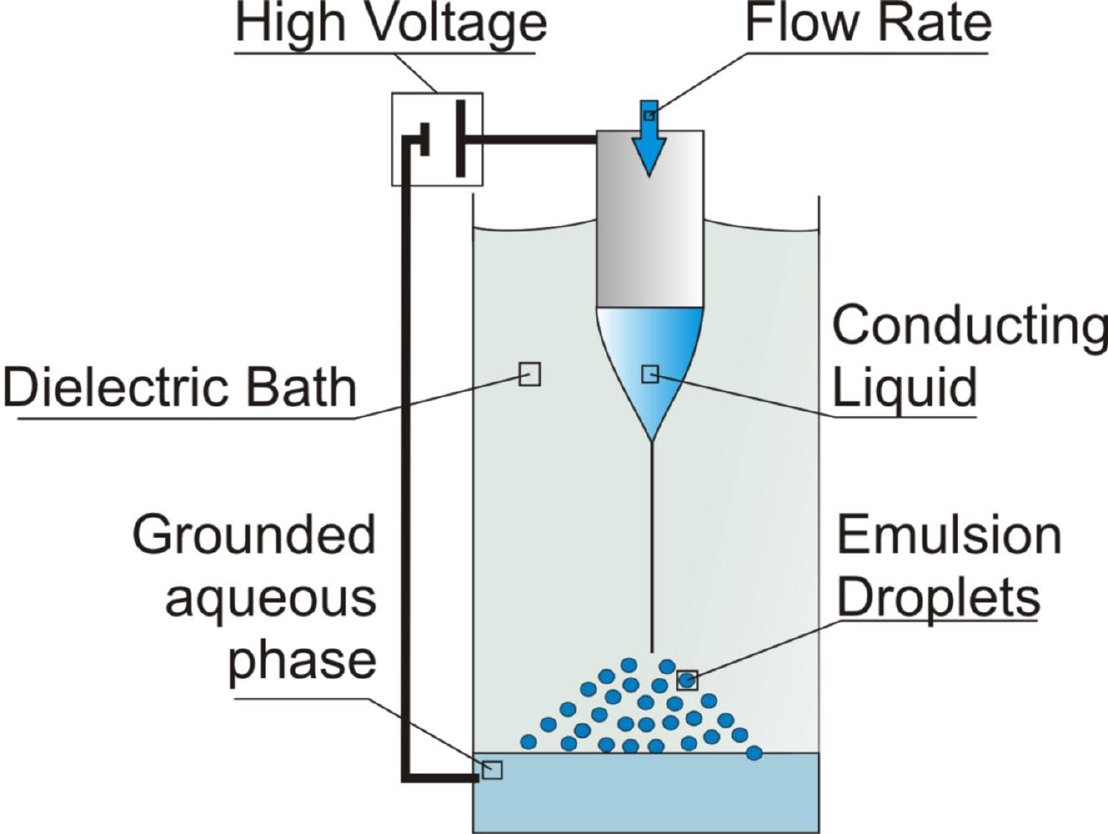}

\caption{Sketch of the experimental set up employed \label{fig:setup}}

\end{figure}

The experiments were carried out in a small cell (see fig. \ref{fig:setup}) made of plexiglass with glass windows to permit visualization. The cell was open in the upper part to insert the injection needle, whose diameter ranged from 0.8/0.6 mm (O.D./I.D.), for the less conductive liquids, and down to 340/20 $\mu m$ with silica capillaries for the most conductive ones. The electrical field is generated by applying a high voltage drop from the injection needle (working as upper electrode) and an electrode placed in the bottom of the cell. Different types of bottom electrodes were chosen depending on the characteristics of the generated hydrosol, more details will be given in the next sections. The liquid was forced through the capillaries using either a syringe pump (Harvard Instruments) or compressed air, depending on the viscosity and conductivity of the sample. Higher conductivity samples require higher control on their flow rates, which can be as small as some nanoliters per hour (\ref{eq:Qo}).

A laser diffraction system (Helos-Sympatec) was employed to perform droplet diameter measurements. In this method a laser beam passes through the hydrosol, the diffraction pattern obtained in this process is collected on a highly sensible multi-element detector. With this patterns as an input, the system deconvolutes the particle size distribution using the Fraunhofer or the Mie solutions for light scattering in spherical particles \cite{mie}. In spite of being a static optical method, the system has been tested and contrasted with many different methods with optimum results.

In table 1 we show the main physical properties of the liquids employed. On the first section of the paper, the liquids employed were glycerine as conductive liquid to disperse and hexane, as host dielectric liquid of the bath. All the experiments were also repeated with heptane, with no appreciable differences on the results. Different samples of glycerine with different conductivities were employed ranging from $10^{-6} S/m$ to almost $10^{-2} S/m$. These conductivities were achieved by adding different types of salt to the glycerine samples (basically NaCl or ClH). 
The inner (glycerine) and outer (hexane, Hheptane) liquids were mainly chosen for their viscosities since we will base our studies to cases in which the viscosity ratio $\lambda=\mu_i/\mu_o\gg1$ (where $\mu_i$ and $\mu_o$ are respectively the viscosities of the inner and outer liquid) which is also the case of atomization in air. In the last section \ref{sec:inverted}, we will briefly discuss some of the effects of the viscosity ratio inversion on the atomization process. 

For the electrical current measurements, the injecting needle was connected indirectly to the positive pole of a high voltage power supply, and different ampmeters were employed to measure the current supplied through the Taylor cone depending on the sample: for the highest currents (close to 1 $\mu$A for the G2 sample), a simple voltmeter with resolution of 1nA was employed. While for the lowest currents (of the order of 1 nA for sample G6) a Keithley picoampmeter was used instead.

Controlling the flow rate is a mayor issue on the these experiments. It is well known that the electrostatic pressure can modify the real flow rate circulating through the meniscus, and for this reason we employed high hydrodynamic resistance lines with small in line diameters. In order to force the liquid through the lines, compressed air was used in most of the cases except for the lowest conducting one, for which a syringe pump (Harvard Instruments) was then employed, since it requires the highest flow rates in these series of experiments (of order of several milliliters per hour).

\begin{table}[!hbp]
\caption{Liquid data\label{table:1}}
\begin{tabular}{|c| c| c| c| c| c| c|}
\hline
& $\rho$(g/ml)& $\mu(mPa \times s)$& $\beta$ & $\gamma$ (mN/m)& I.T.(mN/m)\\

\hline
Glycerol & 1.261        & 1100 &             42.5 &              64  &                       \\

Hexane & 0.655  &       0.294   &           1.9 &                   30 & \raisebox{1.5ex}{28.3} \\
\hline
\end{tabular}
\end{table}

\begin{table}[!hbp]
\caption{Conducting liquid data\label{table:2}}
\begin{tabular}[c]{|c|c|c|c|c|c|c|}
\hline
Sample & G6 & G4(1)& G4(2)&G3(1)&G3(2)&G2\\
\hline
$K(S/m)$& $1,84 \times10^{-6}$&$3,30\times 10^{-4}$& $4,29\times 10^{-4}$& $2,00 \times10^{-3}$& $4,90 \times10^{-3}$& $9.70 \times10^{-3}$\\
\hline
\end{tabular}
\end{table}

\begin{figure}[h]
\centering

\includegraphics[width=0.5\columnwidth]{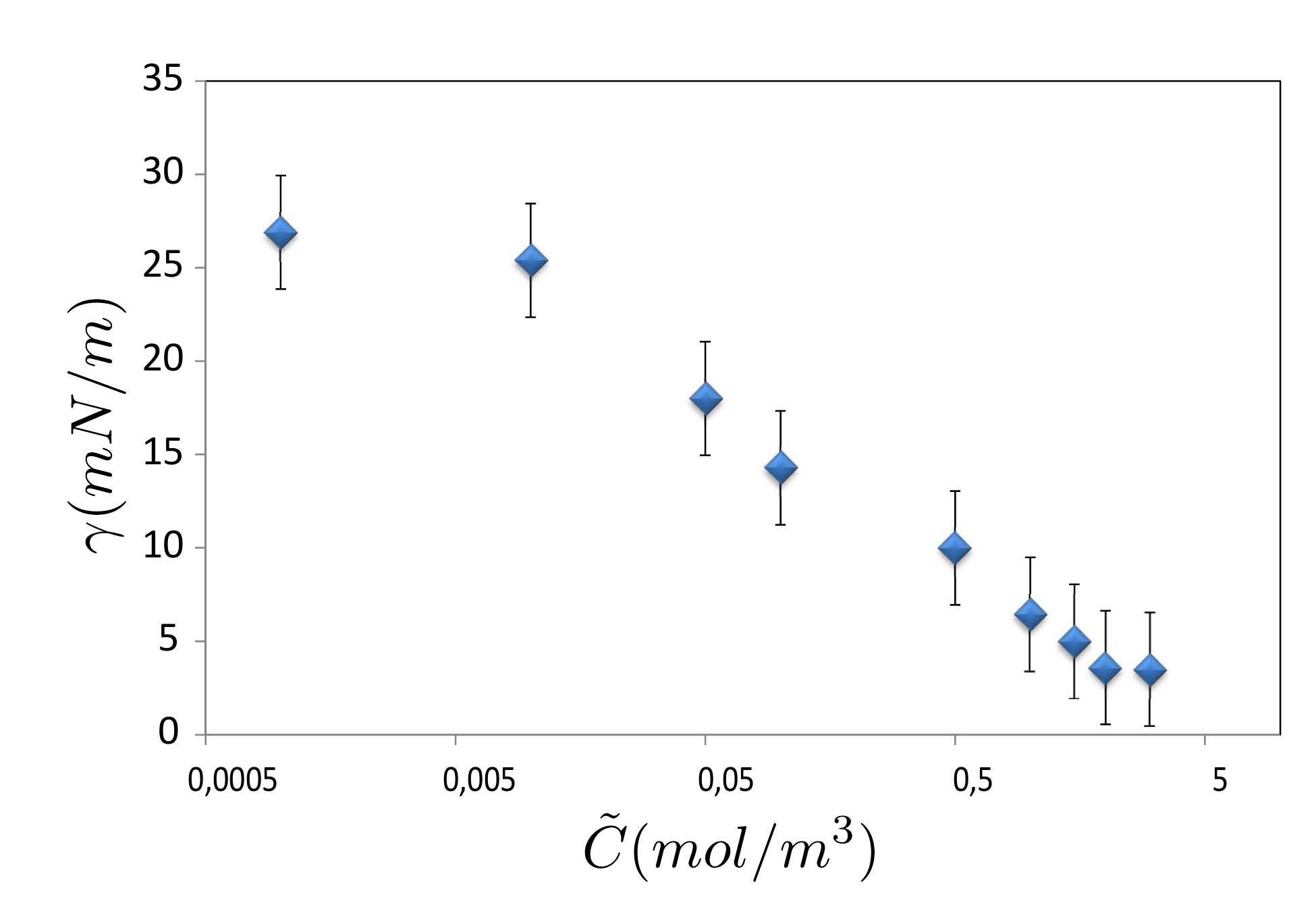}

\caption{Measured interfacial tension for glycerol/hexane against surfactant concentration (span80)\label{fig:cmc}}

\end{figure}

Special care was taken in the measurements of the liquid-liquid interfacial tension with surfactants.  In our case, the surfactant has been added only to the external dielectric liquid (continuous phase), in order to preserve the properties of the inner conductive liquid (dispersed phase). For this purpose, Span 80 (Sorbitan monooleate) was employed in all these series of experiments. In order to find the appropriate surfactant concentration range, surface tension was measured for different concentrations using two different techniques: the Wilhemy plate (Kruss) and pendant drop measurements (KSV Instruments), the results are shown in figure \ref{fig:cmc}.  As can be seen in the plot, surface tension decreases slowly until it approaches a critical concentration when the graph becomes more stepped and reaches it minimum value. This value is reached approximately at $C_{crit}=1.1~mol/m^3$, which we will call our Critical Concentration, i.e. when the value of the surface tension reaches its minimum. Since we are not interested in studying colloidal properties of the surfactant in these processes, we will avoid using the term Critical Micellar Concentration, but just Critical Concentration instead defined as stated above. From now on, all surfactant concentrations $\tilde{C}(mol/m^3)$ will be made dimensionless with the Critical Concentration, so that $C=\tilde{C}/C_{crit}$.

\section{Clean baths}\label{sec:clean}

Barrero et al. \cite{Barr:2003} confirmed that immersed eletrosprays in clean (i.e. surfactant-free), insulating and inviscid liquid baths behave basically as their counterparts in air. Here, we perform extended measurements that will serve us as a reference for the following ones. Before adding a surfactant to any of the fluids or alter the viscosities of the liquids we will confirm by electrical current measurements that the electrospray is certainly being operated in the cone-jet mode. The liquids employed have been chosen in order to resemble the conditions of an electroatomization in air, i.e. a viscous liquid being dispersed in an inviscid fluid ($\lambda=\mu_{i}/ \mu_{o}\gg1$). In this case, we will disperse different samples of glycerol ($\mu_i\approx1Pa\times s$) in hexane ($\mu_o\approx10^{-6}Pa\times s$). The case of inverted viscosity will be briefly described at the end of the paper.

\begin{figure}[hbp!]

\includegraphics[width=0.6\columnwidth]{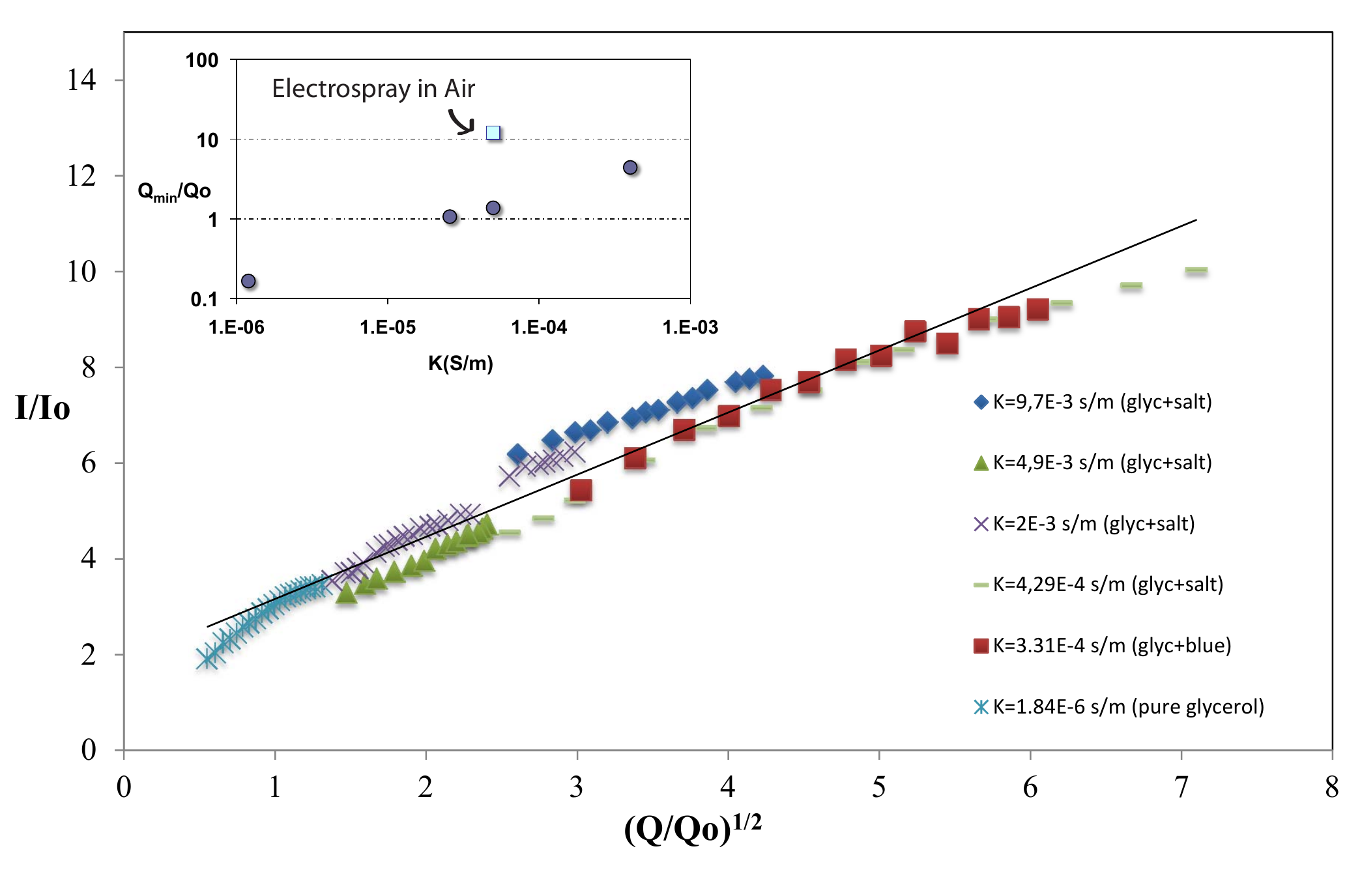}

\caption{Electrical current against flow rate for glycerol samples in hexane\label{fig:IQ}. Insert: Minimum Flow rates for glycerol in hexane. Full circles correspond to the minimum flow rates of immersed electrosprays, the square corresponds to a measurement of the minimum flow rate of an electrospray in air for comparison.\label{fig:qmin}}

\end{figure}

As can be observed in figure \ref{fig:es}, the aspect of an electrospray of this type is completely analogous to those in air. Additionally, the electrical current measurements plotted in figure \ref{fig:IQ} show that most of the samples follow reasonably well a classical square root law of the type in equation (\ref{eq:Ilaw}). This indicates us that the transition from a cone to a jet results from the so-called charge relaxation effects \cite{Loscermora}, i.e. surface charges can not maintain the static cone anymore and they are convected by the liquid flow towards the cone apex, breaking the electrostatic equilibrium of the cone. The measurements fit an equation of the type (\ref{eq:Ilaw}) with an slope $a=1.3$ with less than 10\% of error.

The minimum flow rate was also explored for some of the samples employed. According to several results in the literature\cite{Loscermora, MoraReview, higuera2004current}, the dimensionless minimum flow rate $Q_{min}/Q_{o}$ should only depend on the relative dielectric constant of the liquids. Since we always used the same couple of liquids in the series, the minimum flow rate should be constant for all the samples. However, the results differs from that prediction, and as we can see in the insert of figure \ref{fig:qmin} the minimum flow rate decreases with the conductivity of the liquid, and with values well below those predicted. This result is also found by Bon Ki Ku et al. \cite{Ku:2001} for glycerol samples in vacuum, which make us think that it is a characteristic behavior of glycerol and the nature of the outer medium plays no role. However, their values where in general much higher than ours. 
We encounter here one of the main advantages of the use of electrosprays in liquid mediums: the value of the minimum is flow rate is at least one order of magnitude smaller when the liquid is dispersed in liquid dielectric mediums like hexane instead of air. As an example, the samples with conductivity of order $10^{-5} S/m$ have a minimum flow rate in air of around $400 \mu l/hr$, while for the same liquid in hexane we can reduce the flow rate to $20 \mu l/hr$. The reason can be found in both, the reduction of surface tension (which decreases $Q_o$) and the reduction of the dielectric constant ratio (which decreases $Q_{min}/Q_{o}=f(\beta)$). This has important consequences since we can therefore obtain much smaller droplet diameters by simply dispersing the liquid into a dielectric medium instead of in air. Another advantage is the reduction of the needed voltage and current to work the electrospray: while this particular regime shown in figure \ref{fig:es} requires in air around 6 kV and around 100 nA, immersed in clean hexane it requires circa 3 kV and a few nanoamperes.

%
%
%
%

\begin{figure}[h]
\centering

\includegraphics[width=0.8\columnwidth]{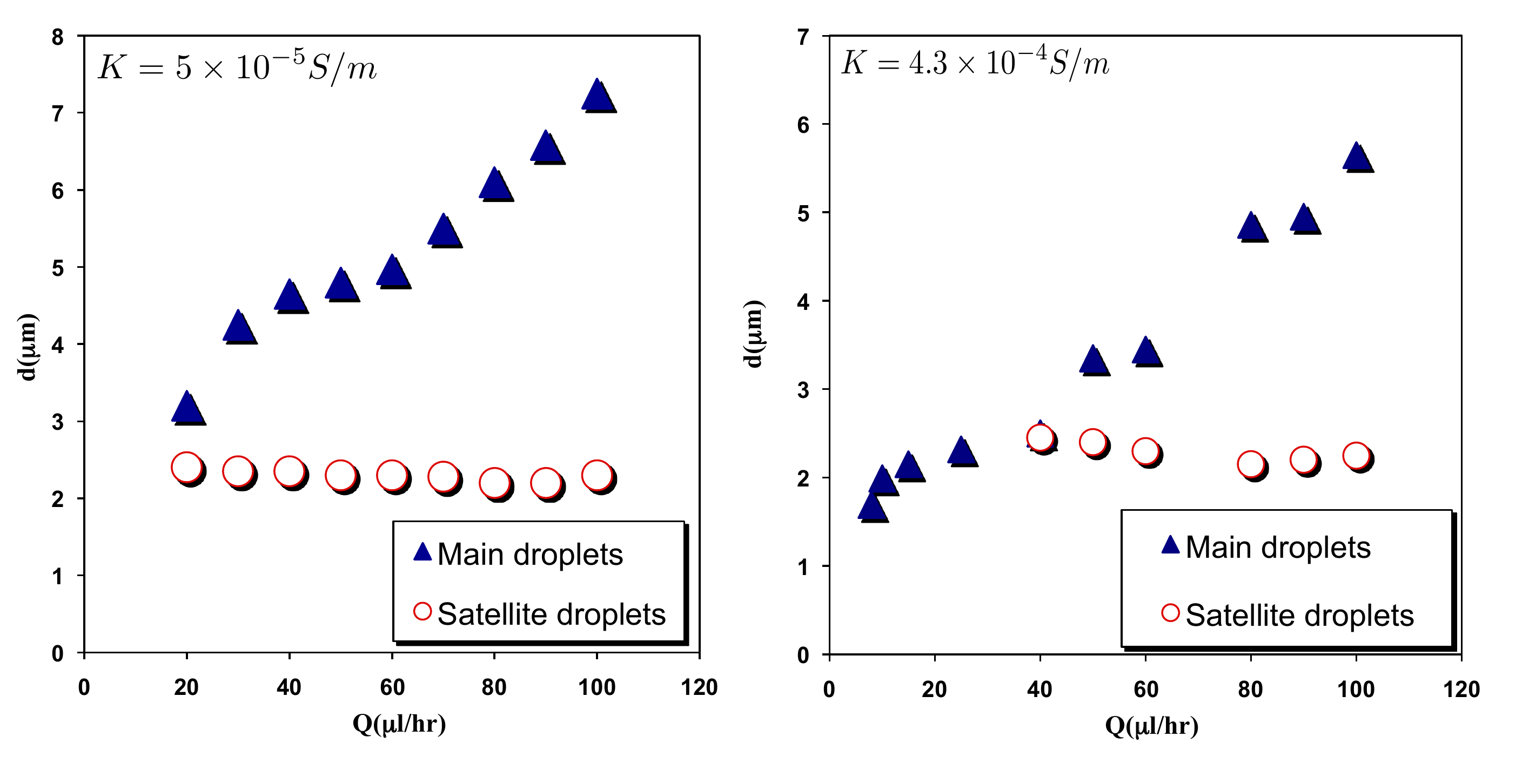}

\caption{Bidisperse droplet distributions for samples G5(1) (left) and G4 (right).  \label{fig:G4G5}}

\end{figure}

\begin{figure}[hbp!]
\centering

\includegraphics[scale=0.8]{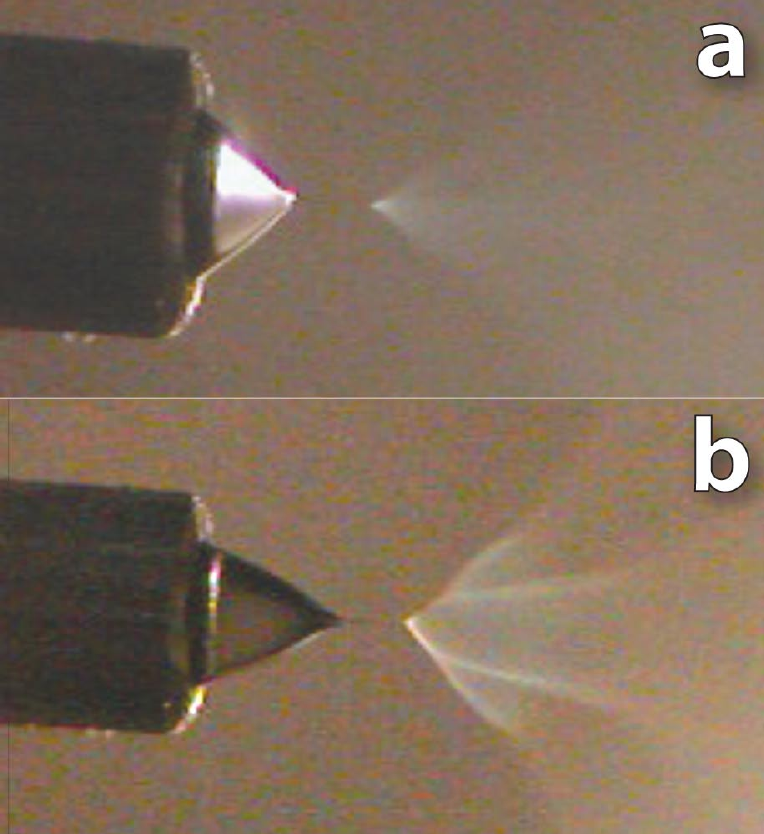}

\caption{Visualization of the hydrosol plume in sample G4. (a) Below the critical flow rate, monodisperse hydrosol. (b) Above the critical flow rate where the droplet size is bidisperse. (enhanced) \label{fig:bidisperse}}

\end{figure}

In order to have precise measurements of the droplet diameters, and since our laser diffraction system only detects particles between 1 and 80 microns, we employed only glycerol samples with conductivities above $10^{-3} S/m$, with diameters ranging from 3 to 45 microns. In table \ref{table:2} we show the samples employed, named by their conductivity as G4 (with $K=10^{-4} S/m$, and so on), G5 and G6.
The droplet distributions of the least conducting samples (as in G5(1) in figure \ref{fig:G4G5}a) are found to be bidisperse for all flow rates , even for the minima observed, being satellite droplets the responsible for the secondary values. The presence of satellite droplets is a well known phenomenon in electrosprays, especially for viscous liquids, and is extensively discussed by Rosell-Llompart and Fern\'andez de la Mora\cite{Rosell}. Although it is natural to have satellite droplets in these regimes, the unsteadiness of the jets enhances it. Certainly, in these samples the jets manifest the so-called ``whipping'' instability\cite{taylor1969electrically, basset} for all flow rates even for the minimum ones. As Riboux et al. recently studied in detail\cite{riboux2011whipping}, the case of the sample G6 is especially impressive: When we operate pure glycerine in air, the jets are perfectly steady and they break-up only by capillary instabilities. However, when inserted in a dielectric liquid bath as hexane, the jet become laterally unstable and bends developing a spiral-like path, reducing its diameter until it breaks by capillary instabilities. Therefore, its distribution of droplets sizes is affected by this violent instability.
However, in the sample G4 we find a different situation (fig \ref{fig:G4G5}b), we observe a monodisperse regime without satellite droplets which is maintained until we increase the flow rate above a critical value in which the distribution turns bidisperse. Both regimes show also very distinctive aspects when observed in operation with the aid of a microscope and special illumination conditions. As shown in figure \ref{fig:bidisperse}, we can observe in the bimodal regime the smaller droplets located in the outer part of the aerosol while the bigger droplets stay in the inner part. This phenomenon was observed already by Zeleny \cite {zeleny} in the first controlled electrospray experiments ever made and explained later by Tang and Gomez \cite {Tang} among others, the reason for this structure in the aerosol can be found in the higher mobility of the satellite droplets, due to their less inertia they are electrically repelled farther in the aerosol. Rosell and de la Mora \cite {Rosell} made an extensive study for several different samples and noted that it was in the most viscous and less conducting ones where such dispersion was more often observed.

\begin{figure}[hbp!]
\centering

\includegraphics[width=0.5\columnwidth]{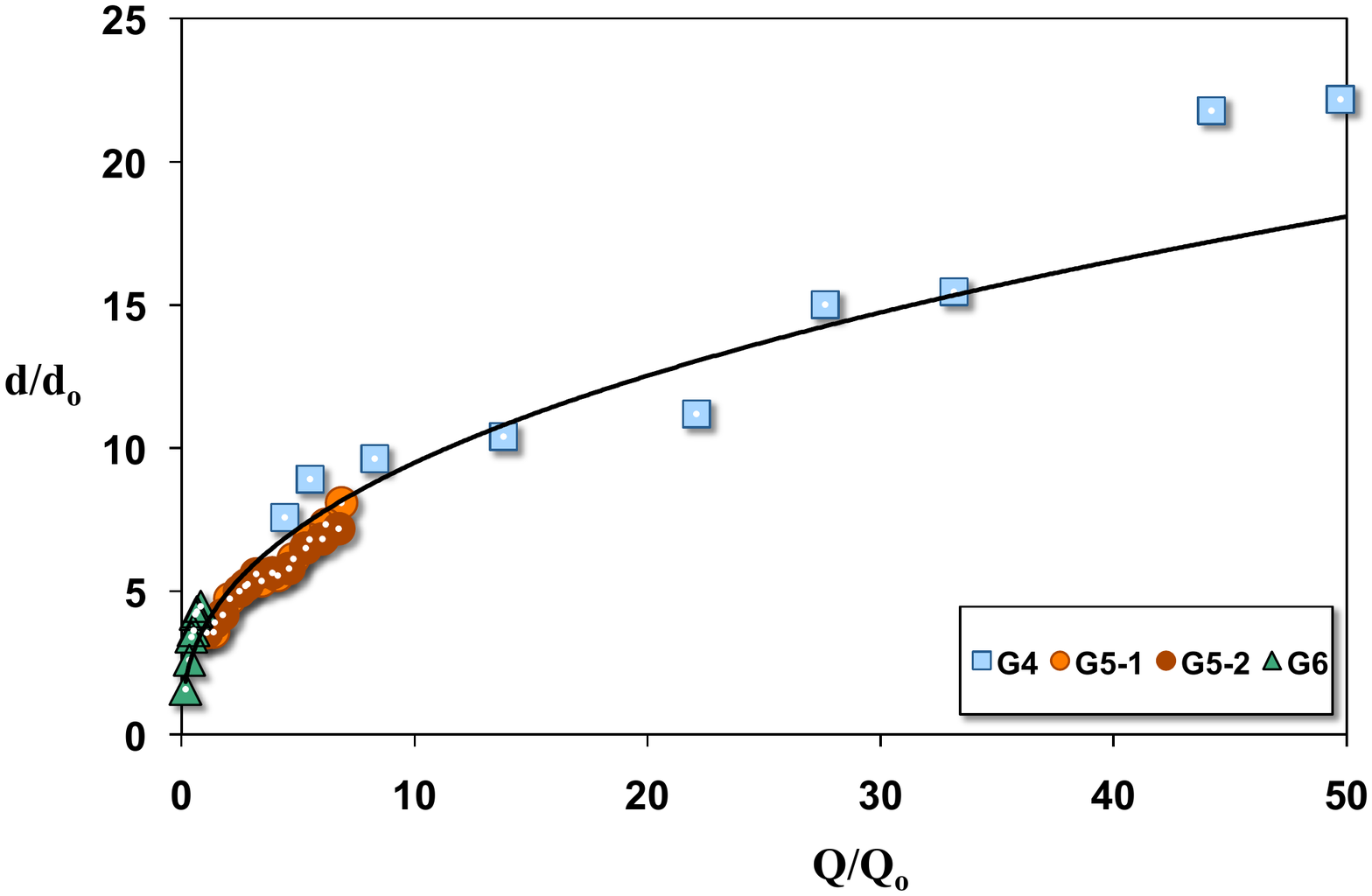}

\caption{Dimensionless representation of droplet size vs injected flow rate. All conductivities sampled follow approximately a $d\sim Q^{1/3}$ law as that described in eq. \ref{eq:dlaw} \label{fig:dQ13}.}

\end{figure}

To analyze the results we will make use of the dimensionless diameter and flow rate defined in equations (\ref{eq:do}) and (\ref{eq:Qo}) respectively. The results are plotted in figure \ref{fig:dQ13}. As we can observed, the data approximately fits a 1/3 law as the one defined in equation (\ref{eq:dlaw}). These results are in accordance with those found by Barrero et al.  \cite{Barr:2003}, although certainly the dispersion of our data also would permit a reasonably good fitting with other similar scaling laws as  $d \sim Q^{1/2}$. According to the recent work of Higuera\cite{higuera2010electrodispersion}, a regime in which $I \sim Q^{1/2}$ and $d \sim Q^{3/8}$ would be a cone-jet mode in which the electric shear stress is all balanced by the inner liquid viscous drag. On the other hand, a result with $I \sim Q^{1/2}$ but a higher exponent for the drop diameter on the flow rate as $d \sim Q^{1/2}$ could be interpreted as a case in which the viscosity of the outer bath becomes dominant. Given the large viscosity ratio in our experiment, and the moderate flow rates employed, we do believe that we are in the first scenario, in which the flow inside the liquid is dominated by the inner viscosity.
It worth noting that all glycerol samples analyzed in hexane presented a very distinctive behavior at higher flow rates, separating from a 1/3 (or 3/8) exponential tendency law to a steeper one (exponents up to 0.578 for sample G6). Many interpretations can be done at this respect:  Rosell-Llompart et al. \cite{Rosell} also reported a similar behavior in their experiments when the critical flow rate for satellite droplets was reached. Certainly in our case, the radial distance from the apex at which the inertial effects become relevant, $R_i=({\rho Q^2/\gamma})^{1/3}$, becomes of the same order as the electrical relaxation distance $r_e=(\beta Q/K)^{1/3}$. This argument ignores however other important effects of the inner or outer viscous stresses\cite{higuera2010electrodispersion} which could be of great importance. We lack however enough evidence to discern among the competing regimes.


\section{Baths with surfactant}\label{sec:surfactant}

In this section we will use a surfactant in the liquid bath (outer phase). In order to make proper comparisons, we keep using the same liquids chosen for the previous chapter. The approach taken in the experiments is the following: first we perform a qualitative exploration varying independently conductivity and surfactant concentration. Once the different regimes are identified in the parameter space constituted by the electrical conductivity K and the surfactant concentration C, we proceed to make electric current and diameter measurements in some selected regimes and discuss the results. The range of the parameters K and C are the following: The conductivity range is the same used in the clean case in the glycerol samples (table \ref{table:1}). The surfactant concentration range was chosen in terms of the value of the interfacial water/glicerine-hexane tension, from its maximum value ($C=0$) to its minimum, achieved at the Critical Concentration ($C=1$). Due to the large amount of data and regimes found, we will illustrate two main extreme cases first with surfactant at its critical concentration value. A lot of intermediate cases have been also found for different values of the parameters $(K, C)$. However, they all share similar characteristics with the two limiting cases shown here and no new phenomena are really introduced. More detailed information about these regimes can be found in the doctoral thesis of A. G. Marin \cite{Alvi}.

\subsection{High Conductivity Cases\label{sec:high}}

\begin{figure}[h]
\centering

\includegraphics[width=0.35\columnwidth]{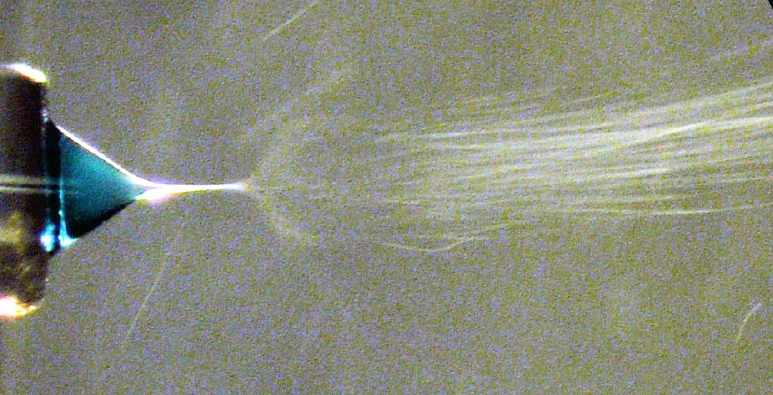}

\caption{Visualization of a high K and high C case with regular long exposures (1ms). \label{fig:blur}}

\end{figure}

To illustrate the extreme case of high conductivity we choose a glycerin sample of $K=4\times10^{-3}S/m$. This conductivity is high enough to illustrate our purposes and is close to the maximum conductivity allowed to make precise measurements and analyze the jet and droplet diameters. In a bath clean of surfactants, this sample gives rise to narrow jets and particles below 1 micron in the low flow rate regime. Adding surfactant in small quantities does not seem in principle to modify the electrospray properties, only a mild decrease in the working voltages. However, when we increase the surfactant concentration above its critical value, what we observe is something completely different. First of all, the voltages needed to obtain the Taylor cone have been reduced by a factor of at least 5 (with an electrode distance of 5 cm, we have 5000 V in a clean bath, while 1000 V in this case). This lower voltage regimes might be of interest with applications in which the state of electrification can be detrimental as in biological solutions. We also note that the flow rate range is much wider. Finally, regarding the meniscus and spray shape, we do not observe the classical conical spray anymore, but only a blurred open spray with an almost cylindrical shape (fig. \ref{fig:blur}).

\begin{figure}[h]
\centering

\includegraphics[scale=0.6]{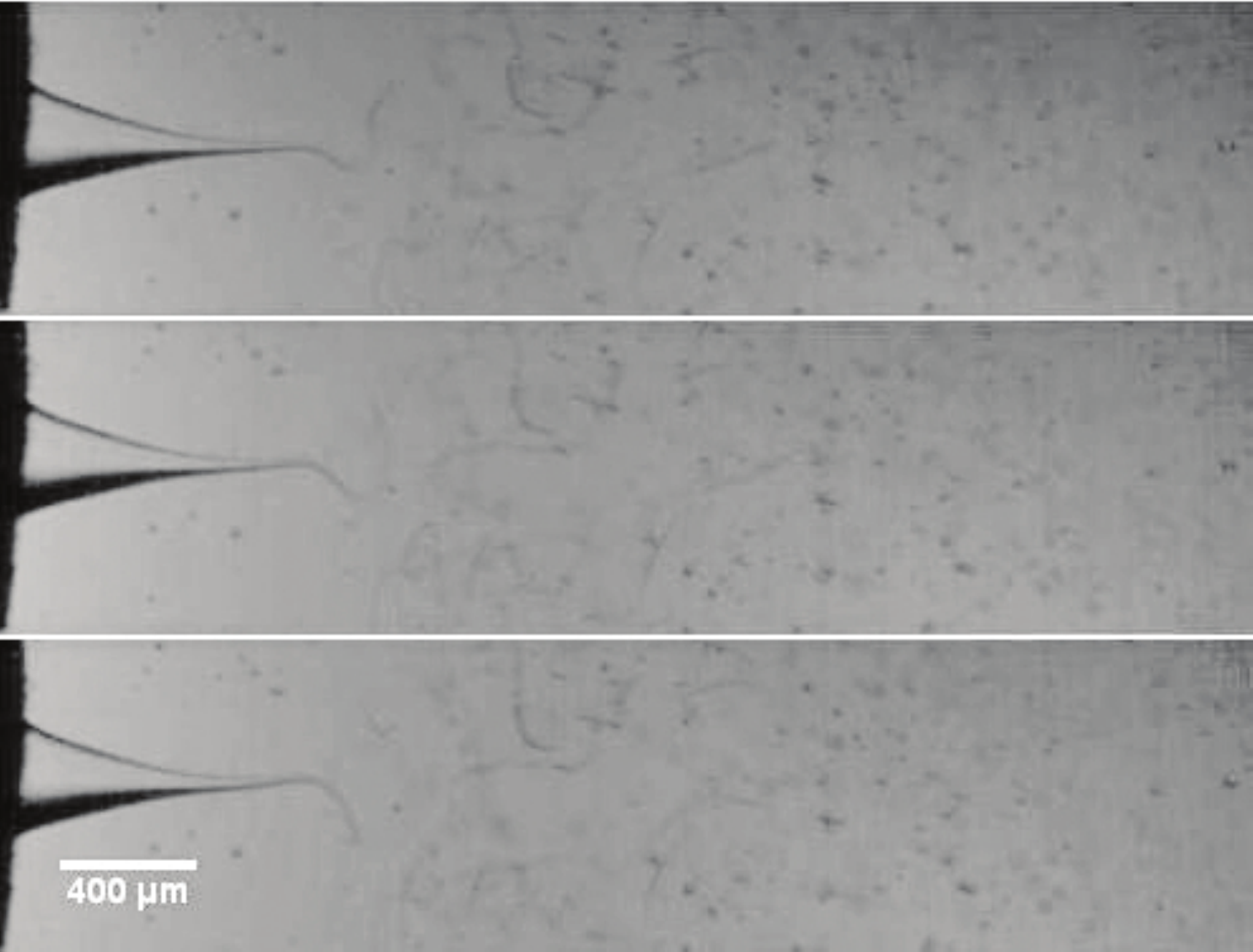}

\caption{Visualization of a high K and high C case with high speed imaging at 2000fps (enhanced) \label{fig:filaments}}

\end{figure}

The reason of this peculiar aerosol shape was revealed using a high speed camera. A sequence of images extracted from a recording are shown in figure \ref{fig:filaments}. As we can see in the images, the jet is much thicker that it was in the clean bath and it soon becomes laterally unstable and breaks immediately in filaments. This contrasts strongly with the classical lateral instabilities, in which the jet performs several ``loops'' before breaking into droplets \cite{riboux2011whipping}.

Due to experimental difficulties it was no possible to use the diffraction system to measure the droplet diameter in these cases where there was a surfactant present in the bath, and therefore we could not collect as much data as in the previous section. The problem found was related to the longer life of the generated droplets within the cell, which made difficult to maintain the bath clean of non-desirable droplets in the region of measurement. A different set-up with a continuous circulating dielectric liquid would solve the problem and would make possible to make better measurements, but as we will discuss later, it is still not clear how the flow in the external liquid would affect the results in this case of extremely low surface tension. However, lot of information can be obtained from high resolution images, high speed videos and electrical current measurements. Depending on the characteristics of the regime, we will proceed with different techniques.

\begin{figure}[hbp!]
\centering

\includegraphics[width=0.6\columnwidth]{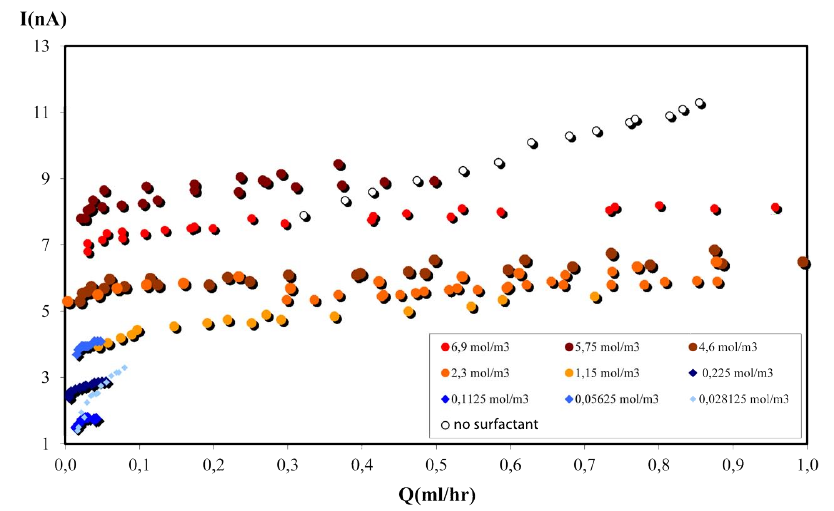}

\caption{Electrical current vs flow rate for different surfactant concentration from the cases of surfactants-free samples (white circles), mild concentrations ($C<1$)(bluish circles) to the high concentrations ($C>1$)(reddish circles).\label{fig:IQspan1}}

\end{figure}

In section \ref{sec:clean}, where high conductive glycerine in clean baths was analyzed, the current to flow rate dependence showed a reasonable good agreement with the well known $I\sim Q^{1/2}$ scaling laws. In this case we will employ a single glycerine sample of relatively high conductivity ($K=6\times10^{-3}S/m$), and let the surfactant concentration ratio vary in two orders of magnitude, below and above the critical concentration. As we mentioned above, there is a critical surfactant concentration in which the regime of atomization changes completely from a regular electrified-capillary breakup to a regime dominated by wider jets and lateral instabilities, yielding highly polydisperse emulsions. 
In figure \ref{fig:IQspan1} the data is plotted in dimensional form. We first note a highly sensitive reaction of the electrical current on the surfactant concentration (i.e. surface tension). It is also very noticeable how the range of stable flow rates for the samples $C<1$ are dramatically decreased due to the reduced surface tension value, as was argued in section \ref{sec:surfactant}, specially when this is compared with the case of dispersion in air. The different dependance on the flow rate of the different samples is also remarkable: almost all $C<1$ samples seem to follow a similar square root law (with different offsets that can be explained by their different surface tensions), while those with $C>1$ present a very reduced dependence with the injected flow rate. Also notice how the addition of surfactant reduces the flow rates and therefore the currents initially, but as the concentration increases the output current surprisingly increases as well; this differs strongly with the classical models of charge transport developed for regular electrosprays and therefore indicates that there must be a different mechanism driving the current in these cases.

\begin{figure}[hbp!]
\centering

\includegraphics[width=0.6\columnwidth]{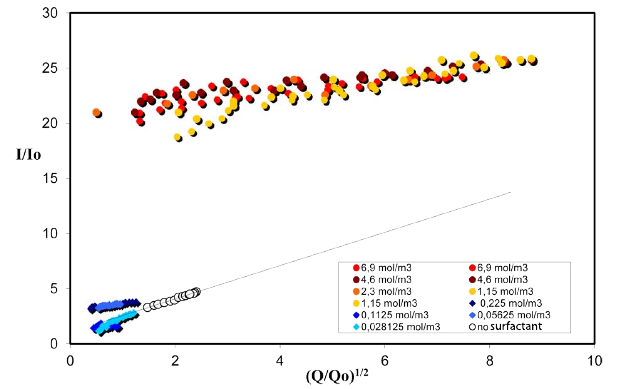}

\caption{Dimensionless Electrical current vs flow rate for different surfactant concentrations, from the cases of surfactants-free samples (white circles), mild concentrations ($C<1$)(bluish circles) to the high concentrations ($C>1$)(reddish circles).  \label{fig:IQspan2}}

\end{figure}

These aspects can be observed in more detail in the dimensionless plot, in which the characteristics flow rate and the current defined in expressions (\ref{eq:Qo}) and (\ref{eq:Io}) has been used. Especial care has been taken with the interfacial tension measurements (see section \ref{sec:setup}). First thing to note is that samples with $C<1$ fit again reasonably well with the well-known square root laws, with some differences for those cases closer to the critical concentration. But most important and surprising is the higher concentration cases, those samples yield electrical currents much higher than those expected for charge relaxation mechanisms \cite{Loscermora}. In the case that the current would be driven by convection, electrical current would be expected to decrease with surface tension since the characteristic velocities needed to maintain a stable jet are being decreased. This is observed for the samples of low surfactant concentration, but not for those with high surfactant concentration ($C>1$) in which the current increases as the surface tension decreases. Additionally, they show a much reduced dependency with the injected flow rate; in the best cases the current increases only a 20\% when the flow rate is increased in one order of magnitude. These features might indicate an electrical current transport for high surfactant concentration samples different than the classical convection-driven mechanisms. 
Unfortunately, the methods employed in previous chapters to obtain droplet diameter distributions were not suitable for the regimes with presence of surfactants. The presence of surfactants increased the stability of the generated droplets, which survived much longer within the liquid bath (they were no absorbed as easily in the water phase at the bottom of the cell) and hindered the laser diffraction measurements. On the other hand, regimes as the one depicted in figure \ref{fig:filaments}, gave rise to a very disordered breakup with wide droplet distributions that could not give much information. For these reasons, high speed video techniques will be used instead with enough resolution to make jet diameter measurements for different surfactant concentration and flow rates, fixing the liquid conductivity in $K=3.1\times10^{-4} S/m$  (higher conductivity would yield jets with diameters that could not be measured accurately, but the conclusions can be extrapolated to higher conductivities).

\begin{figure}[hbp!]
\centering

\includegraphics[scale=0.4]{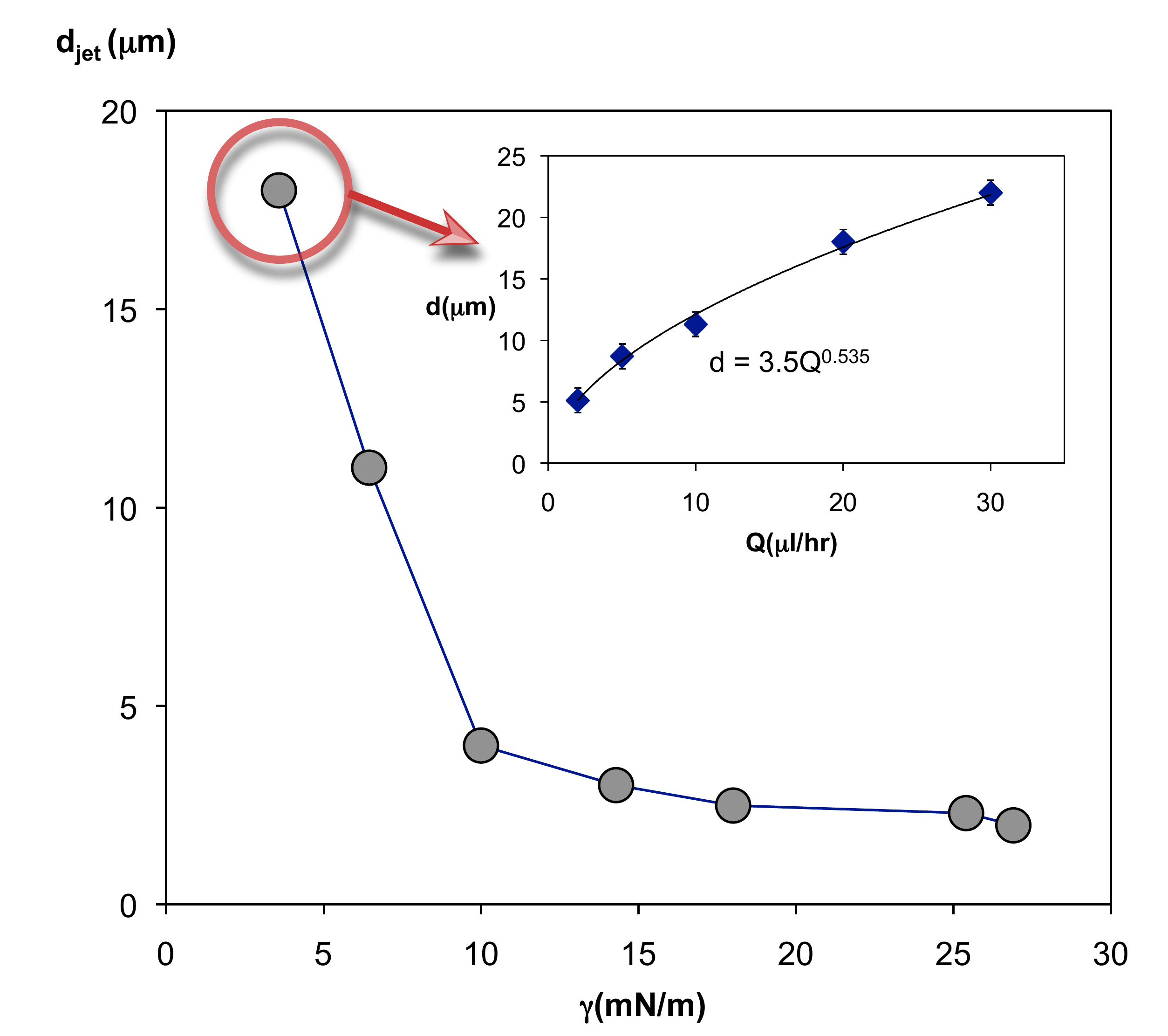}

\caption{Jet diameter variation with interfacial tension for sample of conductivity $K=3\times10^{-4}S/m$
 at flow rate $Q=20 \mu l/hr$ \label{fig:dvsgamma}. Insert: jet diameter is plotted against flow rate, following a scaling law $d\sim Q^{0.53}$ for a sample saturated with surfactant (minimum $\gamma$). }

\end{figure}

Considering classical arguments \cite{Ganan97}, one should expect a decrease in jet diameter as surface tension is decreased as a consequence of the decrease of the characteristic flow rate (see expressions (\ref{eq:do}) and (\ref{eq:Qo})). On the contrary, on figure \ref{fig:dvsgamma} we observe how the diameter increases as surface tension is being decreased (for a given conductivity and fixed flow rate). The increase is timid in the first stages and becomes critical as be reach surfactant concentration values close to saturation.

Regarding the data of $C>1$, not only they showed a reduced dependance with the flow rate, but the also showed a high dependance with the electrical voltage. This effect has not been analyzed throughly, but it has several analogies with the regimes found independently by Gundabala et al.\cite{gundabala2010current} and by Larriba and de la Mora \cite{larriba2011production}. In the first case, Gundabala et al.\cite{gundabala2010current} used a microfluidic co-flow system to produce an immersed electrospray, with relatively low surface tension (no surfactants). They observed regimes in which the current almost does not depend on the flow rate but it depends strongly on the voltage applied. Their conclusion is that, due to the reduced size of the micro-capillary tip, the electrification of the cone-to-jet region was drastically reduced and the electrospray was driven mainly by conduction. The case of Larriba and de la Mora is very different, but the results are analogous. They constructed a coaxial electrospray consisting on an ionic liquid (EMI-BF$_4$) injecting charge micro/nano-droplets into a insulating liquid menisci of heptane (also decane). It is indeed a smart and relatively simple way of atomizing insulating liquids by electrohydrodynamic forces in a controlled fashion. The authors study the atomization regimes of EMI-BF$_4$ by measuring the electric current against the flow rate and the applied voltage in a quiescent bath. They find a very pronounced dependance of the current with the voltage, and a much weaker one with the injected flow rate. Contrary to Gundabala et al. \cite{gundabala2010current}, the authors explain such a behavior by the appearance of space-charge effects of the charged cloud of droplets in the quiescent insulating bath, which would screen the electric field felt by the meniscus.
 
The regimes found by both groups of authors strongly differ with the classical cone-jet electrosprays typically found for conducting, inviscid liquids atomized in air, in which the electrical current depends strongly and almost exclusively on the flow rate due to the convective dominance of the charge transport.  
It would be tempting to identify some of these regimes found in immersed electrosprays with our results, but their conditions do not apply to ours. Regarding the experiments of Gundabala, the smallest meniscus to jet ratio was on the order of 20, which is much higher than their case. On the other hand, we also reject the hypothesis of the space-charge effects in our experiments for two main reasons: First of all, such effects would have been noticed already in clean baths more pronouncedly, but the fact that the current-flow rate dependence showed a clear $I \sim Q^{1/2}$ discards the hypothesis. Secondly, if space-charge effects are important, then the electrical mobility $Z$ of the droplets must relatively low (Larriba and de la Mora found values of $Z\sim10^{-7}$). We can estimate the droplet electric mobility as $Z=2Id^2/9\mu Q$, which yields values of the order of $Z\sim 10^{-5}$ in the samples with $C>1$ (figure \ref{fig:filaments}), comparable to the cases in air, probably due to the relatively large drop size and electric current. 

A more likely explanation for such a regime can be found in the recent study by Higuera \cite{higuera2010electrodispersion}. In this theorical-numerical study of immersed electrosprays, a regime is introduced in which most of the electrical shear stress is equilibrated by the external viscous stresses, which has been normally ignored in the past. Such regime would take place when the dimensional parameters $\Pi=\mu_eK^{1/3}/(\rho_i\epsilon_o\gamma^2)^{1/3}\sim O(1)$ or $\Pi R^{1/4}=\lambda(\mu_i^3 K^2 Q /\gamma^3 \epsilon_o^2)^{1/4}\sim O(1)$. In this particular case of high conductivity and $C>1$, the dimensionless parameters take the experimental values of $\Pi \simeq 0.75\sim O(1)$  and $R\simeq7.75$, and therefore $\Pi R^{1/4}\simeq 1.22 \sim O(1)$. From which we can interpret that, besides the large viscosity ratio, the electrical shear stress on the jet is mostly invested in equilibrate the external viscous shear. This is experimentally supported by the large velocities induced in the external liquid bath. In this regime, a critical flow rate  $Q_M=\epsilon_o\gamma^2 a /\mu_o^2K$ is found, where $a$ is the meniscus radius, and $\mu_o$ the outer liquid viscosity. According to this study, when $Q>Q_M$ the charge transfer region becomes large compared to the meniscus size, the surface charge is therefore limited and the jets are no longer charged-stabilized. In addition, the current becomes independent on the flow rate, and the jet size scales with the flow rate as $r_j \sim Q^{1/2}$. This situation fits better with our data of lowest surface tension due to the strong dependance of $Q_M$ on the surface tension. A decrease of almost one order of magnitude in surface tension in the case of $C>1$ yields $Q_M\simeq 0.6 ml/h$ which is not so far from the range we operate the electropray. To give additional support such hypothesis, we measured the jet diameters for several flow rates for the case of lowest $\gamma$, revealing an exponent of $0.535$, close enough to the predicted. Other symptoms observed here fit with the regime described by Higuera\cite{higuera2010electrodispersion}, specially those regarding the lack of stability of the jets due to the reduced surface charge and the elongated shape of the meniscus (\ref{fig:filaments}). The reason for such critical behavior with the surfactant concentration $C$ above its critical value can be then explained by the fast decrease of the critical flow rate $Q_M=f(\gamma)$ with the interfacial tension $\gamma$ below the working flow rates. 

We should conclude this section by adding some comments on the stability of those regimes with higher surfactant concentration. Certainly in these cases the regimes lack the robustness of the classical cone-jet electrosprays, either in air or in liquid, manifested by oscillations in the electrical current and even in the jet diameter. We will point out two possible reasons for this fluctuations: If we assume that the surfactant concentration is not homogeneous along the liquid interface, it is then very likely that this distribution will be also changing in time depending on the amount of surfactant available in the bath. The liquid in the bath recirculates due to the electrically-driven movement in the conical interface, so there is no control on the amount of surfactant contained in the liquid surrounding the cone-jet.
Related to such effect, a curious phenomenon was observed close to the water phase. This was placed at the bottom of the bath and was used both as ground electrode and as a droplet collector in order to maintain the bath clean of charged particles. However, with surfactant present in the hexane bath, the water-hexane interface was soon covered by surfactant and the coalescence of the charged droplets into the water phase was hindered. As a consequence, it could be observed how some droplets approached the water-hexane interface, ``touched'' it, and they were immediately repelled back to the bath. It seems to be a similar phenomenon as that found by Ristenpart et al. \cite{ristenpart2009non}: a charged droplet migrates towards an oppositely charged interface and, in certain circumstances, instead of being trapped and absorbed, it is repelled from the interface after a short contact. The authors demonstrate that the time required to equilibrate the charge is much shorter than the typical coalescence time; therefore, before the coalescence process is finished, the droplet is already set at the same potential as the interface and a repulsive electrical force is created among them. In our particular case the coalescence time can be also further delayed by the presence of surfactants. Although interesting, the presence of charged droplets surrounding the electrified meniscus can be extremely disturbing and can be also responsible of electrical current and jet diameter oscillations. It is nonetheless a significant problem and  can be easily solved using a continuous external liquid flow as is the case in Gundabala et al.\cite{gundabala2010current} and others \cite{ecoflow,japos}.


\subsection{Low conductivity cases\label{sec:low}}

\begin{figure}[hbp!]
\centering

\includegraphics[scale=0.6]{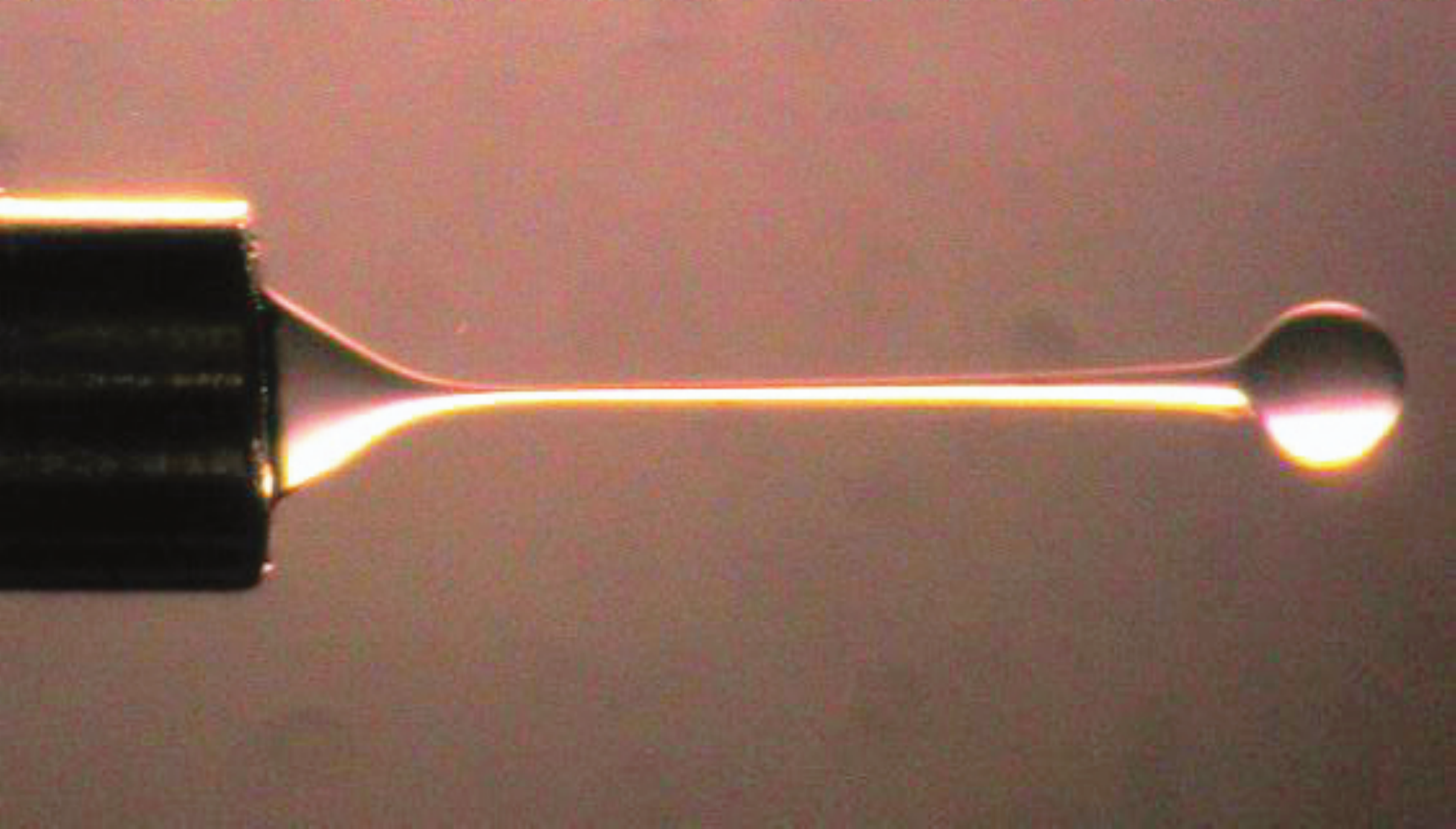}

\caption{Visualization of the ``dripping-jet'' in the limit of low K and high C (enhanced). \label{fig:dripping}}

\end{figure}

We analyze now a extreme case of low conductivity with surfactant concentration above the critical. We use here the least conducting glycerol sample: G6, pure glycerol. It should be reminded that pure glycerine in a clean bath of hexane presents long laterally unstable jets with diameters ranging from 10 to 30 microns depending on the flow rate. Adding small quantities of surfactant gives a similar result as in the high conductivity case, lower working voltages and certain increase in the jets size. But as we reach the critical concentration, we find a surprising situation like the one showed in figure \ref{fig:dripping}.

This mode requires only 20\% of the voltage needed to stabilize an immersed electrospray in a clean bath, but instead of having a thin laterally unstable jet, we find a broad widening jet that will end up growing a droplet in its end. This droplet increases its size as the jet elongates until the electrical forces are strong enough to detach it from the jet tip, then the jet retracts and a new droplet starts to grow again. The size of the droplet is still small enough to ignore gravity effects: a Bond number defined as $Bo=\rho_i g R_c/\gamma$, where $\rho_i$ is the liquid density,$\gamma$ its surface tension, and $R_c$ its typical drop size, yields values of $Bo\sim0.01$. Therefore the detached droplets are directly driven to the electrode surface. Although the electrical current was extremely low, it was possible to make some spurious measurements of oscillating electrical currents varying with the same frequency as the dripping frequency.
A relatively thick jet ($60-100 \mu m$) generates and widens in its tip until a big droplet is created ($200-400 \mu m$). Since new liquid is being injected, the droplet increases its diameter as the jet becomes longer and slender until a certain point in which the droplet reaches such a size that detaches from the jet. The liquid ligament then retracts, a new small droplet is formed on its tip, and the process starts again. The frequency of this cycle, lengths and diameter of the jet and size of the droplet may vary depending on the controlled parameters: voltage and flow rate. An increase on voltage will make longer jets, and therefore (for the same flow rate) smaller droplets, with a slight increase in the frequency. A reduction in flow rate will create shorter jets, smaller droplets and a higher frequency. Regarding the electrical current, it was mostly too low to be measured accurately, but it presented fluctuations synchronized to the drop generation cycles.
The key of this behavior seems to be in the joint effect of reduced surface tension and low conductivity in the liquid. Certainly, the electrical field needed to create a conical structure in the meniscus is so reduced that it fades away not so far from the meniscus tip and consequently the liquid jet flows downstream practically as a free jet, feeling no electrical stress in its surface. This differs from regular Taylor cones in air, where the electrical field close to the tip is so strong that the jet suffers a tremendous electrical stress still far from the cone. The electrical stress is responsible of the progressive decrease in size of electrified driven jets and also explains their resistance against varicose instabilities \cite{PepeHerrera2004,Tomotika}. In our case, the absence of this electrical stress explains the progressive widening of the jet. Additionally, the reduced electrical field on the jet also explains the absence of lateral instabilities, often related to jets electrically overcharged.
However, the arguments above still can not explain the appearance of a droplet in the tip of the jet. On the other hand, a similar phenomenon has been observed and successfully explained in co-flow systems to generate monodisperse emulsions \cite{AlbertoLC}. In that case, Utada et al.\cite{Utada1} observed how the droplet to jet diameter ratio increased its value greatly when the inner liquid jet was being slowed down by the outer slower liquid, the authors identified this phenomenon as the outcome of an absolute instability due to the decrease in inertia of the jet, which triggers the classical transition to a ``dripping'' regime in the jet at certain distance from the needle\cite{clanet1999transition}.
A classical example of absolute capillary instabilities can be found in the dripping at the end of a tube, which is characterized by the growth of perturbations in fixed points of the interface. Monodisperse droplets are therefore produced due to the proximity of the injection needle which damps most of the external perturbations. On the other hand, in the case of convective instabilities the maximum growth rate of the perturbations is found far from the injection needles, giving rise to polydisperse droplets due to the effect of external perturbations. In our experiment, the electrical field is enough to generate an electrified conical meniscus through which the liquid in being injected at certain flow rate. Within the meniscus, the fluid is being accelerated towards the vertex until inertia overcomes the capillary tension a jet is formed. The jet is then accelerated by the electrical field of the meniscus, however, due to the low value of both electrical field and surface charge, the tangential stress over the surface decreases fast with the distance from the tip. At this point, the only force over the surface of the jet is the viscous drag from the outer liquid which is not in any way negligible, but we have no control of it, and therefore its only effect is to slow the jet down, in a similar way as it does the outer liquid in Utada et al. \cite{Utada2} experiments. The jet then widens up and velocity decreases until inertia can not maintain a stable jet anymore and finally a droplet appears at its tip. In co-flow experiments, a series of oscillations in the interface preceded the detachment of the droplet from the tip; we are not aware of these oscillations in our experiments, but they may be visible employing higher spatial or temporal resolution techniques. Nevertheless, it must be noted that the liquid interface in this case presents certain surface charge (although low when compared to that found in classical electrosprays, but still not negligible), which seems to give the interface some resistance to varicose oscillations\cite{Basaranjets}. This is of course an extreme case of low conductivity and therefore wide jets, but the same phenomenon has been found increasing the conductivity of glycerine with thinner jets, giving rise to the same type of instability but with smaller drops.

\section{Notes on inverted viscosity ratios}\label{sec:inverted}

So far in this article we have experimentally studied immersed electrosprays of highly viscous liquids in quiescent and inviscid liquid baths with variable surface tension, controlled by the use of surfactants. A very different case that deserves study is that of the atomization with inverted viscosity ratio, i.e. an inviscid liquid being atomized in a viscous insulating liquid. We will refer this regimes as $\lambda=\mu_i/\mu_o<1$, where $\lambda$ is the viscosity ratio, and $\mu_i$ and $\mu_o$ are respectively the inner and outer viscosities. This is the typical case, for example, of  water being electrosprayed in a bath of oil. Barrero et al. \cite{Barr:2003} reported successful experiments of water in oil, but they were not extensively studied. Unfortunately we can not provide enough data about this case of inverted viscosities, but we can very briefly discuss the preliminary experiments performed, specially regarding the stability of the process. Water samples of different conductivities were atomized in a bath of silicone oil (viscosity of 20 cP) for different surfactant concentrations. For the case of clean baths, the electrospray was hardly ever stable and the current showed a high dependance on the voltage applied. In the case with surfactants in the viscous bath, stability was achieved, but still with a high sensitivity to the applied voltage. A significant symptom observed was the extremely reduced velocity of the emitted droplets, which were easily observed with at standard recording speeds. 
More data is obviously needed, but the symptoms observed seem to point out the presence of strong space-charge effects as those described by Larriba and de la Mora \cite{larriba2011production}. If this is the case, the presence of surfactants might have improved the stability due to the increase of droplet size, which increases the droplet electrical mobility and therefore reduces the screening of the electrical field.  We conclude that, according to our experiments, space-charge effects would become relevant when the viscosity of the outer liquid is similar or higher than that of the atomized liquid, although further research must confirm such hypothesis.

\section{Summary and conclusions}\label{sec:summary}

In this work, we have reviewed the behavior of immersed electrosprays of viscous conducting liquids into insulating inviscid liquids baths, for different parametric cases, but mainly focused in the effect of the surface tension modified by the presence of surfactants. As a summary, we have confirmed the results by  Barrero et al.
\cite{Barr:2003} regarding the reproducibility of cone-jet regimes in immersed electrosprays. We also make emphasis in the advantages of the immersed electrosprays, since the moderate reduction of surface tension brings lower characteristic flow rates and therefore a lower droplet size. However, a more pronounced reduction of surface tension, achieved by the use of surfactants, provokes the loss of the cone-jet regime, characterized for its reproducibility and monodisperse sprays. On the other hand, the use of surfactant also brings advantages as the reduction of the voltage necessary for atomization (of interest when cells or vesicles are encapsulated inside the conducting liquid), and higher droplet electrical mobility. The new modes of atomization at minimum surface tension have been analyzed and compared with different new regimes recently identified by Higuera\cite{higuera2010electrodispersion} and Larriba and de la Mora\cite{larriba2011production}. The nonessential inconveniences originated from the use of surfactants, as the surfactant accumulation at the base of the meniscus, or the enhanced stability of the hydrosol, could be easily overcome by a controlled external flow, in a similar fashion as Gundabala et al. \cite{gundabala2010current} did in their experiments. 
Other effects need to be further studied, such as the effect of the outer liquid flow on the stability of the regimes, specially in liquids of similar or higher viscosity than the conducting liquid. The control of the flow in the outer liquid could also solve the hypothetical space-charge effects that could appear when the viscosity ratio both liquids have similar viscosities. However, the effect of the external viscous shear stress on the surface of the Taylor cone have been only marginally studied and would be of special relevance in the case of reduced surface tension by surfactants, when the electrification of the meniscus is also the lowest. To conclude, we do believe that electrohydrodynamic atomization has several features that could be of great interest for lab-on-chip applications which has to be still explored. Therefore, the aim of this work has been to study possible solutions to the current problems that the community has been facing in recent years.

\begin{acknowledgements}
A. G. M. and I. G. L. dedicate this work to Prof. Antonio Barrero Ripoll, respectively mentor and colleague of the authors, who inspired this and many other research along his life. A. G. M. owes a late acknowledgement to the Yflow crew: J. E. D\'iaz, M. Lallave and D. Gal\'an, for their help and their friendship. A. G. M. also acknowledges the encouragement of Prof. Juan F. de la Mora to finally publish these results.
\end{acknowledgements}

\bibliographystyle{unsrt}

\bibliography{bibelectrospray.bib}

\end{document}